\documentclass[11pt]{article}
\usepackage{moriond,epsfig}

\def\Journal#1#2#3#4{{#1} {\bf #2}, #3 (#4)}

\def\NPB{{\em Nucl. Phys.} B}

\def\PRL{\em Phys. Rev. Lett.}
\def\PRD{{\em Phys. Rev.} D}

\def\be{\begin{equation}}
\def\ee{\end{equation}}
\def\bea{\begin{eqnarray}}
\def\eea{\end{eqnarray}}

\begin{document}
\vspace*{4cm}
\title{NON-SUSY SEARCHES AT THE TEVATRON}

\author{ J. M. HAYS, for the CDF and D\O\ collaborations. }

\address{Department of Physics, High Energy Physics Group, Imperial College London, \\ London SW7 2BZ, United Kingdom}

\maketitle\abstracts{As the integrated luminosity of Run II at the Tevatron increases, the sensitivity to new physics is approaching and in some cases surpassing that previously attained in Run I. Presented below are a selection of new results from searches for physics beyond the Standard Model at the CDF and D\O\ experiments. }

\section{Introduction}
CDF and D\O\ continue to search for physics beyond the Standard Model at the upgraded Tevatron in Run II. The increase in centre-of-mass energy to 1.96 TeV is of particular importance for searches for high-mass objects where significant increases in production cross-section are expected. Results from a sample of analyses from both experiments are presented here, utilising between 40 pb$^{-1}$ and 75 pb$^{-1}$.

\section{Leptoquark searches}
Both CDF and D\O\ are conducting searches for first and second generation leptoquark pair production, assuming no cross generational couplings and a 100\% branching fraction to electron or muon plus quark. This process gives rise to an experimental signature of two leptons plus two jets and no missing transverse energy with a dominant background of $\gamma^*/Z\rightarrow ll + $ 2 jets.
 Presented here are results from the first generation search at CDF and the second generation search at D\O.
For the first generation search events are used with two central electrons with transverse energy $E_T > 25$ GeV and two jets where the leading jet has $E_T > 30$ GeV and the next highest $E_T$ jet has $E_T > 15$ GeV. The invariant mass of the dielectron pair is required to be outside the range  76 GeV $< M_{ee} < 110$ GeV. CDF observes no excess of events in 72 pb$^{-1}$ and sets a limit on the leptoquark mass of $M_{LQ} > 230$ GeV at 95\% confidence level which compares favourably with the Run I limit of 213 GeV \cite{cdflq}. The D\O\ limit on this channel from Run I was 225 GeV \cite{d0lq}.
D\O\ require events to have two opposite signed muons above 15 GeV with two jets above 20 GeV and a dimuon mass $M_{\mu\mu} > 110$ GeV. Utilizing around 40 pb$^{-1}$ of available data no signal events are observed and a mass limit is set at $M_{LQ} > 157$ GeV.

\section{Searches for high-mass objects}	
	
\subsection{Large extra dimensions}
D\O\ searches for large extra dimensions, assuming the Standard Model particles are confined to a 3-brane with gravity propagating in the extra dimensions. 
This gives rise to a signature of an excess of high mass dilepton and diphoton events from the couplings to Kaluza-Klein (KK) gravitons. The contribution of the gravitons can be characterised by the parameter $\eta_G$ in equation \ref{eq:led1} where $M$ is the dilepton or diphoton invariant mass, $\theta^*$ is the scattering angle in the dilepton or diphoton rest frame and $f_{SM}$, $f_{KK}$ and $f_{\mathrm{inter}}$ represent the Standard Model, KK-graviton and interference terms in the cross section respectively.

\begin{equation}
	\frac{d^2\sigma}{dM\cos{\theta^*}} = f_{SM} + f_{\mathrm{inter}}\eta_G + f_{KK}\eta^2_G
\label{eq:led1}
\end{equation}
Events are selected in the dimuon channel by requiring two muons matched to central tracks with a transverse momentum $p_T > 15$ GeV	 and a dimuon mass $M_{\mu\mu} > 40$ GeV. The dielectron and diphoton channels are combined by not requiring a matched track to reconstructed electromagnetic (EM) objects in the calorimeter. The event selection requires two EM objects with $E_T > 25$ GeV and the missing $E_T < 25 $ GeV. Backgrounds are estimated using fast Monte-Carlo for dilepton and diphoton production with instrumental and reconstruction fake rates being estimated from the data. Fits to data and background in the $(M, |\cos{\theta^*}|)$ plane show no excess of events. This can be used to set an upper limit on $\eta_G$ which translates into a lower limit on the fundamental Planck scale. Results for different formalisms \cite{formalism1,formalism2,formalism3} are shown in table \ref{tab:led1}. The limit using the di-EM channel approaches the D\O\ Run I result while the dimuon analysis was not performed in Run I and so is a new limit.
\vspace{-0.5cm}
\begin{table}[hb]
\caption{ Lower limits on the fundamental Planck Scale in TeV \label{tab:led1}}
\vspace{0.4cm}
\begin{center}
\begin{tabular}{|l|c|c|c|c|} \hline
	Formalism & GRW & HLZ, n=2 & HLZ, n=7 &Hewett, $\lambda = +1$ \\ \hline
        Di-EM     & 1.12& 1.16     & 0.89     & 1.00 \\
        dimuon    & 0.79& 0.68     & 0.63     & 0.71 \\ \hline   
\end{tabular}	
\end{center}
\end{table}
\vspace{-0.5cm}
\subsection{Randall-Sundrum Gravitons}
\begin{figure}
\begin{center}	
\begin{tabular}{rr}
\hfill \epsfig{figure=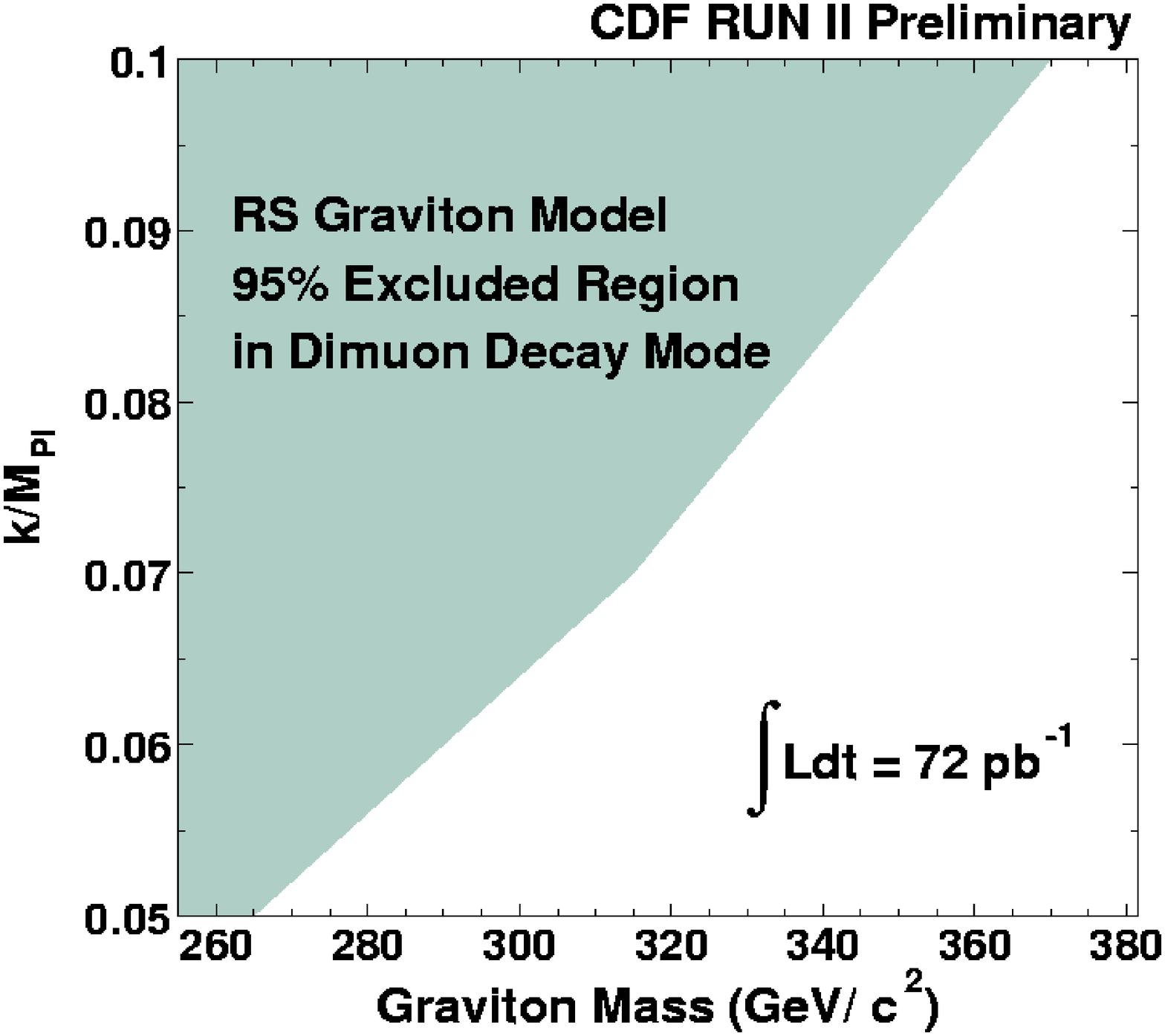,height=2.0in} & \hfill \epsfig{figure=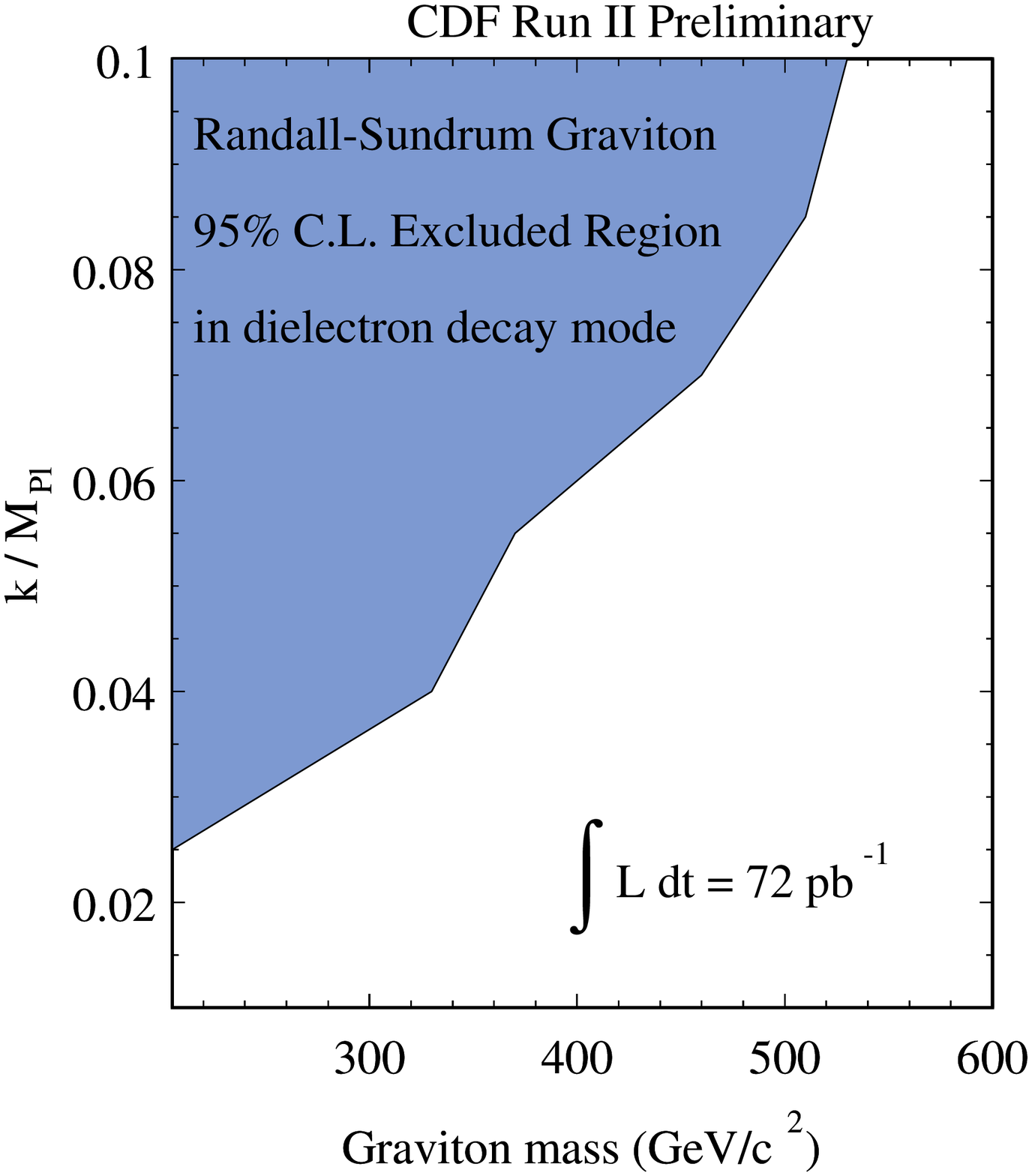, height=2.0in}\\\end{tabular}	
\end{center}
\caption{Randall-Sundrum graviton exclusion regions in the dimuon and dielectron decay modes\label{fig:rs}}	
\end{figure}

Looking at dilepton events, CDF searchs for an excess of high mass objects above the Standard Model backgrounds arising from the production of excited RS-gravitons in 5 dimensions. No excess is observed and so a region is excluded at 95\% confidence level in the plane of the free parameters: graviton mass $M_G$ and coupling $k/M_{PL}$ as shown in figure \ref{fig:rs}. These results are competitive with those from LEP.
\subsection{Z$^\prime$ to dilepton}
CDF and D\O\ look for Z$^\prime$ production in the $qq \rightarrow Z^\prime \rightarrow e^+e^-$ channel with CDF peforming an additional search in the dimuon channel. In the dielectron channel CDF requires two electrons to be well reconstructed with $E_T > 25$ GeV where at least one of the electrons must be in the central region and the event missing $ E_T / \Sigma E_T < 2.5$. D\O\ selects events with two electrons each having $E_T > 25$ GeV. For the dimuon channel events are selected with two muons, each with $p_T > 20$ GeV. Neither experiment observes an excess of events above the Standard Model backgrounds. In the dielectron channel D\O\ sets a limit, using 50 pb$^{-1}$, of $M_{Z^\prime} > 620 $ GeV which approaches the Run I result of $M_{Z^\prime} > 670$ GeV \cite{d0zprime}. CDF has set limits on the $Z^\prime$ mass at $M_{Z^\prime} > 650$ GeV and $M_{Z^\prime} > 455$ GeV for the dielectron and dimuon channels respectively. This can be compared with a combined dilepton Run I result of $M_{Z^\prime} > 690$ GeV \cite{cdfzprime}.

\section{Signature based searches}	
In addition to carrying out model specific analyses, both experiments also use a model independant signature based approach.    
\subsection{dijet mass spectrum}
Using their entire inclusive jet sample (75 pb$^{-1}$) CDF select the two highest $E_T$ jets in each event with $|\eta| < 2 $ and $|\tanh{\Delta\eta/2}| < \frac{2}{3}$, where $\eta$ is pseudo-rapiditiy and $\Delta\eta$ is the difference in $\eta$ for the two jets. A simple parameterisation of the Standard Model background is fitted to the dijet invariant mass spectrum to look for \emph{bumps} comparable with the mass resolution. No significant excess is observed beyond the Standard Model background. This allows limits to be set on the production of a variety of new particles, summarised in table \ref{tab:dijet} together with the Run I results \cite{cdfdijet}.
\vspace{-0.5cm}
\begin{table}[htb]
\vspace{0.4cm}
\caption{Excluded mass limits in GeV for new particles decaying to dijets
\label{tab:dijet}}
\begin{center}
\begin{tabular}{|l|c|c|}
\hline
 Particle          & RunI result   & Run II limits \\ \hline 
axigluons          & $200 < M < 980$ & $200 < M < 1130$ \\
excited quarks     & $200 < M < 570$ & $200 < M < 760$ \\
                   & $580 < M < 760$ &               \\
colour octet 
technirhos         & $260 < M < 640$ & $260 < M < 480$ \\
E6 diquarks        & $280 < M < 420$ & $290 < M < 420$ \\
W$^\prime$         & $300 < M < 420$ & $300 < M < 410$ \\ \hline
\end{tabular}
\end{center}	
\end{table}
\vspace{-0.5cm}
\subsection{ e$\mu$ final state analysis}
\begin{figure}
\begin{center}
\begin{tabular}{cc}	
\epsfig{figure=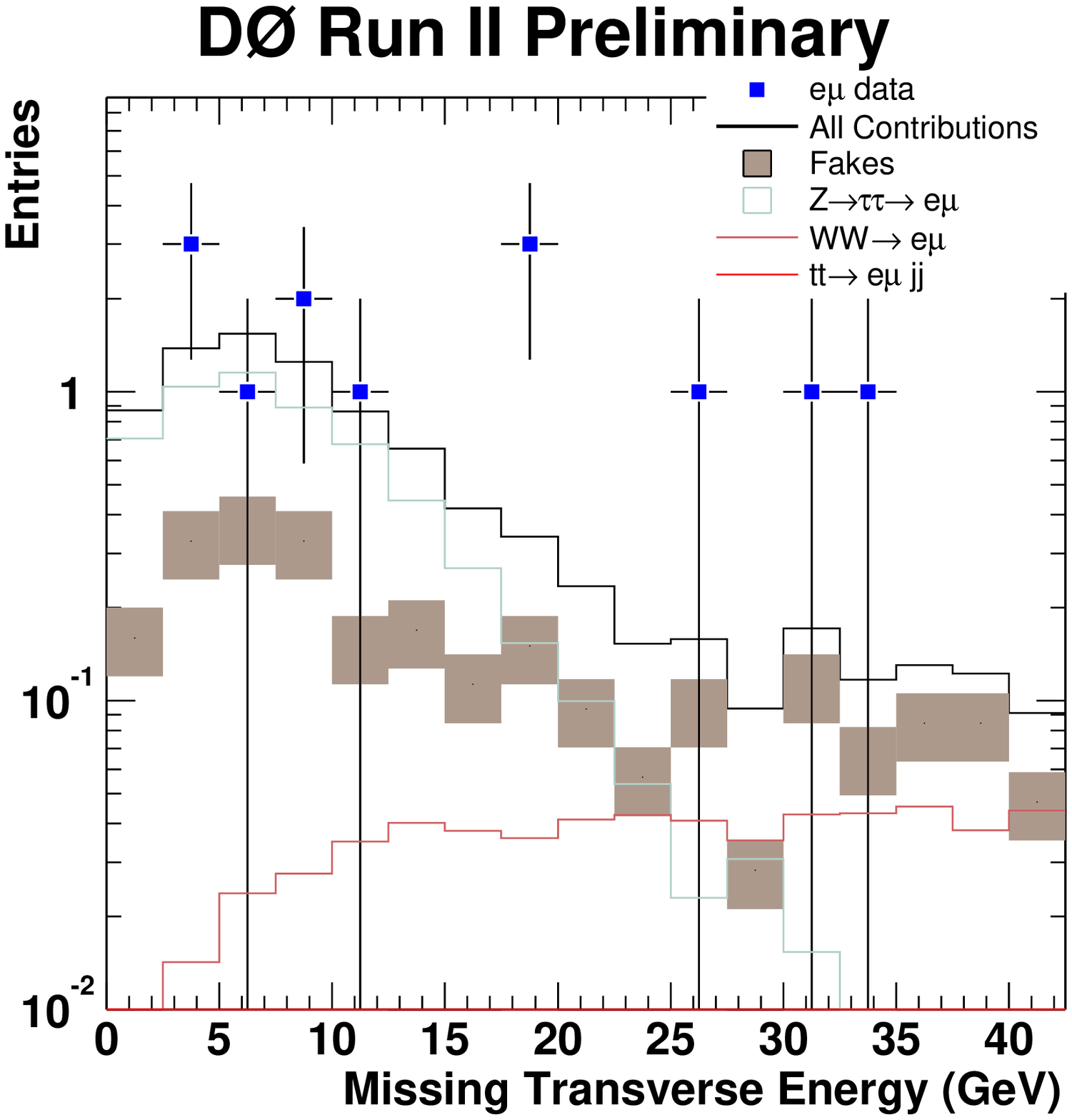,height=2.0in} \hfill

& \epsfig{figure=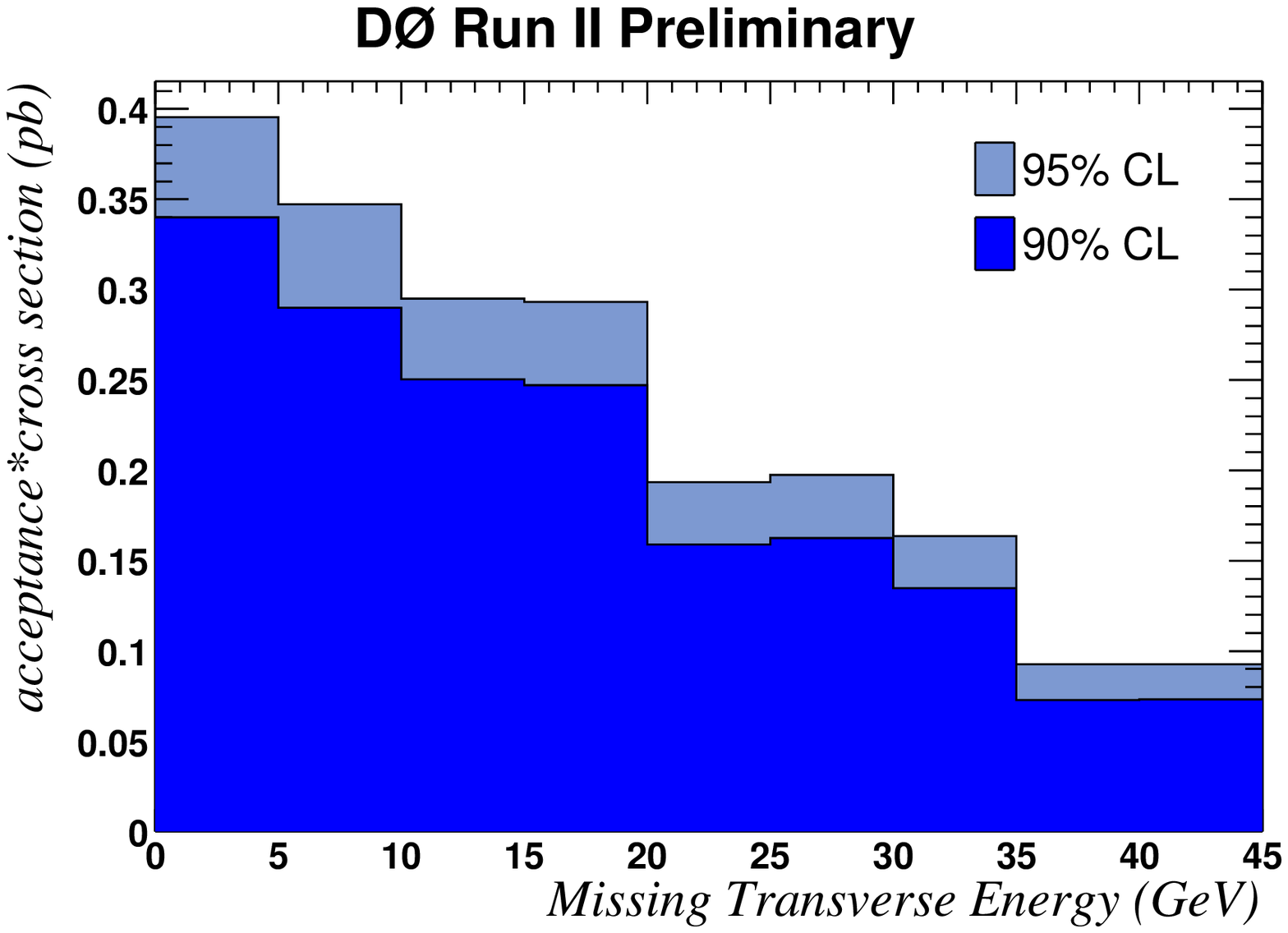, height=2.0in}\\
\end{tabular}
\end{center}	
\caption{e$\mu$ final state data and expected backgrounds and the resulting limit on acceptance $\times$ cross section as a function of missing transverse energy\label{fig:eu1}}
\end{figure}

The very low Standard Model backgrounds allow a model independant approach for this analysis at D\O. Events are selected with at least one electron and at least one muon both with $p_T > 15$ GeV and with no reconstructed jets. The main physics backgrounds come from Z$\rightarrow \tau \tau$ at low missing $E_T < 25$ GeV with WW and $t\bar{t}$ dominating for higher missing $E_T$. The physics backgrounds are estimated from simulation with the fake rates being extracted from the data. The left panel of figure \ref{fig:eu1} shows good agreement between data and backgrounds for 33pb$^{-1}$. This sets 95\% confidence limits on the acceptance $\times$ cross section for new physics as shown in the right panel of figure \ref{fig:eu1}.
\subsection{Charged massive particles}
The upgraded CDF detector in Run II includes a new Time of Flight (TOF) detector which provides a better particle identification than $dE/dx$ for particles with higher $\beta\gamma$. Tracks with a long time of flight are selected from events firing high $p_T$ muon triggers with $p_T > 40$ GeV to ensure full tracking efficiency. A reference time, $t_0$, is established by looking at tracks with low $p_T < 20$ GeV. Events are then selected which contain a track with $TOF - t_0 > 2.5$ ns. 7 events are observed in 53 pb$^{-1}$ of data with the background estimated from tracks in the range $ 20 < p_T < 40$ GeV to be $2.9 \pm 0.7 (\mathrm{stat}) \pm 3.1 (\mathrm{sys})$ events. This can be interpreted in the context of stable stop production to give a limit on the stop mass of $M_{stop} > 107$ GeV at 95\% confidence level.

\section{Conclusions}
CDF and D\O\ continue to actively pursue searches for exotic new phenomena.
Although no new particles or physics have been discovered in Run II, the discovery potential rises as the integrated luminosity accumulates.  Results are now approaching and improving on Run I and with the improved reach of the upgraded detectors and accelerator a new signal could be waiting to be discovered.

\end{document}